\documentstyle[12pt]{article}
\def\be7{\hbox{$^{7}$Be}}

\begin{document}
\def\frontiers{{\it Frontiers of Neutrino Astrophysics}, ed. by Y.
Suzuki and K. Nakamura (Universal Academy Press, Inc., Tokyo, Japan,
1993)}
\def\gtorder{\mathrel{\raise.3ex\hbox{$>$}\mkern-14mu
             \lower0.6ex\hbox{$\sim$}}}

\noindent
{\Large\bf Solar Neutrinos:  What We Have Learned\footnote{Invited Talk,
{\it Physical Processes in Astrophysics}, International Meeting
in Honour of Evry Schatzman, September 1993,
Paris, France}}
\vglue1cm
\noindent
John N. Bahcall
\vglue.5cm
\noindent
Institute for Advanced Study, Olden Lane, Princeton, NJ 08540
\vglue1cm
\noindent
{\bf Abstract.}\
Four solar neutrino experiments are currently taking data. The results of
these experiments confirm the hypothesis that the
energy source for solar luminosity is hydrogen fusion.
However,
the measured rate for each of the four solar neutrino experiments
differs significantly (by factors of 2.0 to 3.5) from the corresponding
theoretical prediction that is
based upon the standard solar model and the simplest
version of the standard electroweak theory (zero-neutrino masses, no flavor
mixing).

\medskip
\noindent
If standard electroweak
theory is correct, the energy spectrum for \b8 neutrinos created in
the solar interior must be the same (to one part in $10^5$) as the known
laboratory \b8 neutrino energy spectrum.
A direct comparison of the chlorine and the
Kamiokande experiments, both of which are sensitive to \b8 neutrinos,
suggests that the discrepancy between theory and
observations depends upon neutrino energy, in conflict with standard
expectations.
Monte Carlo studies with 1000 implementations of the standard
solar model indicate that the chlorine and the Kamiokande experiments
cannot be reconciled unless new weak interaction
physics changes the shape of the \b8 neutrino energy spectrum.
The boundary conditions that the solar model luminosity equals the
current observed photon luminosity and that the solar model must be
consistent with helioseismological measurements are two of the
strongest reasons that the predictions of the standard solar model are robust.

\medskip
\noindent
The results of the two gallium solar neutrino experiments strengthen the
conclusion that new physics is required and help determine a relatively small
allowed region for the MSW neutrino parameters.
New experiments that will start in 1996
will test--independent of solar models--the inference
that physics beyond the standard electroweak model is required to
resolve the solar neutrino problem.
\vglue.5cm
\noindent
\vfil\eject
\baselineskip=18pt
\section{Evry Schatzman and Nuclear Energy\hfil\break
 Generation}

\noindent
Like everyone here,
I am indebted to Evry Schatzman for insight and for inspiration
concerning a
number of different problems that will be discussed in this symposium.
But, I would like to draw special attention to a fundamental
contribution that he made in 1951 which has become an essential element
in the discussion of the solar neutrino problem and of the more general
question of how main sequence stars shine.  In an important paper in
{\it Comtes Rendus}, Schatzman~\cite{Schatzman51} pointed
out that the most likely
termination of the $pp$ chain was via the reaction
$^3He~+~^3He \rightarrow ^4He ~+~p~+~p$.
He correctly stressed that the reaction he was proposing was most likely
at relatively low stellar interior temperatures and at relatively high
densities.
The $^3He-^3He$ reaction involves
the fusion of two ambient $^3$He nuclei and, to the best of my
knowledge, had not been previously discussed. Equation (3) of
Schatzman's paper describes what is currently believed to be
the dominant cycle for the
fusion of four protons to provide energy in the solar interior.
The suggestion of the dominant role of the $^3He~+~^3He$ reaction is
only one of Schatzman's many fundamental contributions but
it illustrates the remarkable originality, depth, and breadth of his
thinking.

\section{Introduction}
\label{sec-intro}

\noindent
I will review the present status of solar neutrino astronomy and
solar neutrino physics, with special emphasis on the discrepancy between the
predicted and the observed counting rates in the experiments designed to
detect solar neutrinos.

Since this symposium is partially a historical retrospective, it is
interesting to begin with an ironic aspect of
the proposal in 1964~\cite{Bahcall64,Davis64}
 that a practical solar neutrino experiment could be carried out
using a chlorine detector.  If you look back at those two papers, you will
see that the only motivation presented for doing the experiment was to use
neutrinos ``...to see into the interior of a star and thus directly
verify the hypothesis of nuclear energy generation in stars.''
The energy-generating process being tested is

\begin{equation}
4 p \longrightarrow ^4He ~+~ 2e^+  ~+~ 2\nu_e ~+~ 25~MeV,
\label{eq:Heburning}
\end{equation}
by which four protons are burned to form an alpha particle, two
positrons, two neutrinos, and thermal energy.

The goal of demonstrating that Eq.~(\ref{eq:Heburning})
is the origin of sunshine
has been achieved.  Solar neutrinos have been
observed in four experiments with, to usual astronomical accuracy
(a factor of two or three), about the right numbers
and about the right energies.
Moreover, the fact that the neutrinos come from the sun was established
directly by the Kamiokande~II experiment which showed that electrons
scattered by neutrinos recoil in the forward direction from the sun.
These experimental results represent,
in my view, a great triumph for the physics, chemistry,  and astronomy
communities since they bring to a successful conclusion the development
(which spanned much of the 20th century) of
a theory of how main sequence stars shine.

However, most of the current interest in solar neutrinos is focused
on an application of solar neutrino research that was not discussed or
even considered at the time of the original experimental and theoretical
proposals.
It has subsequently been realized that one can use solar neutrinos for
studying experimentally aspects of the weak interactions that are not
currently accessible in laboratory experiments.
These studies of new physics are based upon the
quantitative discrepancy between the predictions and the
observations for solar neutrinos.  To evaluate the significance of these
discrepancies, one
must carry out more precise calculations and pay closer attention
to the theoretical uncertainties than is conventional in most stellar interior
studies.  I will therefore discuss at some length the
uncertainties in the theoretical calculations.

Nearly everyone in this room is an astronomer.
Therefore, you will immediately recognize how the possible discovery of
new physics with solar neutrinos differs from the astronomical
discoveries with which you are familiar.  Astronomical discoveries, like
the finding of quasars, of pulsars, of x-ray binaries with neutron stars
or black holes, of strong infrared sources,
of x-ray bursters and $\gamma-$ ray bursters,
of very young stars and very old
galaxies, all resulted from pointing telescopes with exceptional
equipment and finding something unpredicted but recognizable by
qualitative features.
Unfortunately, discoveries made using solar neutrinos are different.
No one has an intuitive feel for how many solar neutrino events ought
(or ought not) to be
seen per year in a large detector.  Precise quantitative
predictions must be made in order to determine if we have learned
something new. The estimated uncertainties in those predictions are
crucial for deciding on whether discoveries have been made.

When we compare solar neutrino calculations with solar neutrino
observations,
we begin with a combined standard model, the standard
model of electroweak theory plus the standard solar model.
We need the standard solar model to tell us how many neutrinos of what
energies are produced in the solar interior.  And, we need the standard
electroweak model--or some modification of the standard electroweak
model--to tell us what happens to the neutrinos after they are created.
We need to know how neutrinos are affected when they pass through the
enormous amount of matter in the sun and travel the great distance from
the solar interior to detectors on earth.

Do neutrinos change their
flavor from electron-type to some other type during their journey from
the sun to the earth?  The simplest version of the standard electroweak
model says: ''No.''  Neutrinos have zero masses in this model and lepton
flavor is conserved.  Nothing happens to the neutrinos after they are
created.

It turns out that one can learn an enormous amount about
neutrinos by observing experimentally what happens to solar neutrinos
after they are created.  This fact is largely responsible for the great
current interest in solar neutrinos.

There are four operating solar neutrino experiments, three of which
use radiochemical detection (one chlorine and two gallium detectors)
 and one detector which is electronic (the Kamiokande pure water
detector).

The first, and for two decades the only, solar neutrino experiment uses
a radiochemical chlorine detector to observe electron-type
neutrinos via the reaction~\cite{Davis64}:

\begin{equation}
\nu_{\rm e} + {\rm~^{37}Cl} \to {\rm e^-} + {\rm~^{37}Ar} .
\label{eq:Clreaction}
\end{equation}
The $^{37}$Ar atoms produced by neutrino capture are extracted
chemically from the 0.6 kilotons of fluid, $C_2Cl_4$, in which they are
created and are then counted using their characteristic radioactivity in small,
gaseous proportional counters.
The threshold energy is 0.8 MeV.
The chlorine solar neutrino experiment is described by
Davis~\cite{Davis93} and references quoted therein.

The second solar neutrino experiment to have been performed, Kamiokande
II~\cite{Hirata89,Hirata91,Suzuki93}
is based upon the
neutrino-electron scattering reaction,

\begin{equation}
\nu + {\rm e} \to \nu^\prime + {\rm e}^\prime,
\label{eq:NuScattering}
\end{equation}
which occurs inside the fiducial mass of 0.68 kilo-tons of
ultra pure water. Only \b8 solar neutrinos are detectable in the
Kamiokande~II experiment, for which the lowest published value for the
detection threshold is 7.5 MeV.
In the Kamiokande~II experiment, the electrons are detected by the Cerenkov
light that they produce while moving through the water.
Neutrino scattering experiments provide information that is not available
from radiochemical detectors, including the direction from which the
neutrinos come, the precise arrival times for individual events,
information about the energy spectrum of the neutrinos, and some
sensitivity to muon and tau neutrinos.

The fact that the neutrinos are coming from the sun is established
 by the Kamiokande~II experiment since the electrons are
scattered in the forward direction in reaction
Eq.~(\ref{eq:NuScattering}).  The observed directions of
the scattered electrons
trace out the position of the sun in the sky.

There are two gallium experiments
in progress, GALLEX~\cite{Anselmann92,Anselmann93} and
SAGE\ \cite{Abazov91a,Abazov91b,Bowles93},
that provide the first observational
information about the low energy neutrinos from the basic proton-proton
reaction. The GALLEX and SAGE
experiments make use of neutrino absorption by gallium,

\begin{equation}
\nu_{\rm e} + {\rm~^{71}Ga} \to {\rm e^-} + {\rm~^{71}Ge} ,
\label{eq:Gareaction}
\end{equation}
which has a threshold of only 0.23 MeV
for the detection of electron-type neutrinos.
This low threshold makes possible the detection of
the low energy neutrinos from the proton-proton (or $pp$) reaction;
the $pp$ reaction initiates the nuclear fusion chain in the sun by producing
neutrinos with a maximum energy of only 0.42 MeV.
Both the GALLEX and the SAGE
experiments use radiochemical procedures to
extract and count a small number of atoms from a large detector,
similar to what is done in the chlorine experiment.

Figure~1 shows a comparison between the predictions of the standard
model~\cite{Bahcall92} and the four operating solar
neutrino
experiments~\cite{Davis93,Hirata89,Hirata91,Suzuki93,Anselmann92,Anselmann93,Abazov91a,Abazov91b,Bowles93}
.  The unit used for the three radiochemical
experiments is a $SNU ~=~ 10^{-36}$ events per target atom per second.
The result for the Kamiokande water experiment is expressed, following
the experimentalists, in terms of a ratio to the predicted event rate.
The errors shown are, in all cases, effective $1\sigma$ uncertainties, where I
have combined quadratically the quoted statistical and systematic errors.
I will use throughout this review the standard solar model results of
Bahcall and Pinsonneault~\cite{Bahcall92} since this is the
only standard solar model published so far
to take account of helium diffusion.  However, accurate solar models
without helium diffusion have been published by many
other authors and are in good agreement with the Bahcall-Pinsonneault
solar model without helium diffusion.

All four of the solar neutrino experiments yield values less than the
predicted value for that detector and outside the combined errors.  I
shall present later in this talk
a detailed comparison between the the theoretical
predictions
and the measured rates.  However, one fact is
apparent already from Figure~1.  The discrepancy between theory and observation
is about a factor of 3.5 for the chlorine experiment, whereas the
discrepancy is only a factor of 2.0 for the Kamiokande experiment.
These two experiments are primarily sensitive to the same neutrino
source, the rare, high-energy  \b8 solar neutrinos (maximum neutrino
energy of 15 MeV).  Thus the disagreement between theory and experiment
seems to depend upon the threshold for neutrino detection,
being larger for chlorine
(0.8 MeV threshold) than for the Kamiokande (water) experiment (7.5 MeV
threshold).  This may be the most significant fact about the solar
neutrino problem.

\section{Theoretical Neutrino Fluxes}
\label{sec-theoretical}

\noindent
Table~1 shows the solar neutrino fluxes computed with the aid of the
standard solar model.  The $pp$ neutrino flux is predicted to be
the largest flux by an order of magnitude, but is not observable in the
chlorine and in the Kamiokande experiments.  Only the gallium
experiments have a low enough threshold to be sensitive to the $pp$
neutrinos.  The second most abundant neutrino source is $^7$Be, which
produces two lines.  The $^7$Be neutrinos are expected to contribute a
small amount to the capture rate in the chlorine experiment (15
\% of the total standard model prediction) and a somewhat larger
fraction (25\% of the total rate) to the gallium experiment, but are
below threshold in the Kamiokande experiment.

\begin{table}
\begin{center}
\begin{tabular}{c@{\hspace{1in}}l}
\multicolumn{2}{c}{Table 1.\ Neutrino Fluxes~\cite{Bahcall92}}\\
\noalign{\medskip\hrule\medskip}
Source&\hfil Flux\hfil\\
\noalign{\medskip\hrule\smallskip\hrule\medskip}
&$(10^{10}~{\rm cm}^{-2}{\rm s}^{-1})$\\
p-p&$6.0~(1 \pm 0.007)$\\
pep&$1.4 \times 10^{-2}~(1 \pm 0.012)$\\
hep&$1.2 \times 10^{-7}$\\
$^7$Be&$4.9 \times 10^{-1}~(1 \pm 0.06)$\\
$^8$B&$5.7 \times 10^{-4}~(1 \pm 0.14)$\\
$^{13}$N&$5 \times 10^{-2}~(1 \pm 0.17)$\\
$^{15}$O&$4 \times 10^{-2}~(1 \pm 0.19)$\\
\noalign{\medskip\hrule}
\end{tabular}
\end{center}
\end{table}

The most easily detected neutrinos are the very rare, but higher-energy,
\b8 neutrinos.  They are predicted to be four orders of magnitude
rarer than the low-energy $pp$ neutrinos, but because the \b8
neutrinos have relatively high energies they dominate the predicted
capture rate for the chlorine experiment (almost 80\% of the total
predicted rate) and are the only neutrino source to which the Kamiokande
experiment is sensitive.

Table~1 shows the most important
neutrino fluxes and the effective $1\sigma$ error bars
that have been calculated with the standard solar model.
The size of the uncertainties is of critical importance.  I have
therefore devoted a full chapter, Chapter~7, in my
book {\it Neutrino Astrophysics}
to the estimation of the errors in each neutrino flux.  For a recent
detailed calculation of the errors and a comparison with the uncertainties
estimated by different authors, see Ref.~\cite{Bahcall92}.

The $pp$ neutrinos are calculated with a precision that is better than
1\%.  The next most abundant neutrinos, the $^7$Be neutrinos, are
calculated with an uncertainty of $\pm 6$\%.  The rare \b8 neutrinos
are calculated with the least accuracy, $\pm 14$\%.  Unfortunately, the
easier solar neutrinos are to detect, the more difficult they are to calculate.

\section{Comparison of the Chlorine and the Electron-Scattering
Experiments with
Theory}
\label{sec-Comparison}

\noindent
We will now compare the results of the
chlorine and  the electron-scattering (Kamiokande)
experiments with the theoretical expectations for each experiment.
The predicted event rate, $8 \pm 1$ SNU, for the chlorine experiment is
dominated by the 6.2 SNU from the rare \b8 neutrinos.  The next most
important source, according to the standard model, for this experiment
is the electron-capture line from $^7$Be, which is predicted to produce
a 1.2 SNU capture rate.  The $pep$ and CNO neutrinos are expected to
produce together a rate of 0.6 SNU.

The experimental rate is~\cite{Davis93} $2.28 \pm 0.23$ SNU.

In order to assess the significance of the disagreement between theory
and observation for the solar neutrino experiments, I have performed a
series of Monte Carlo calculations.  The results are shown in Figure~2,
which was constructed using the results from a thousand implementations
of the standard solar model.  For each model, all of the important
input parameters (including nuclear reaction rates and
chemical composition) were chosen from normal distributions that had
means and standard distributions equal to the experimentally-determined
values.  For a each solar model, every parameter was chosen from its own
normal distribution and the solar calculations were iterated
to match the observed characteristics of the present-day sun.  This
procedure is required in order to take account of the strong effects of
boundary conditions and the coupling of different calculated neutrino
fluxes that exists among the solutions of the coupled partial
differential equations of stellar evolution.

None of the 1,000 solar models represented in Figure~2
has a neutrino flux that is in agreement
with the observed rate.

Figure~3 shows a similar comparison for the neutrino-electron scattering
(Kamiokande~II) experiment and the 1000 solar models.  The Kamiokande~II
experiment is only sensitive to the high-energy side of the \b8 neutrino
energy spectrum.  Although for the Kamiokande~II experiment none of the
1000 solar models are consistent with the observed value, the
discrepancy is only a factor of two~\cite{Hirata91,Suzuki93}
in this case (compared to the factor
of 3.5 for the chlorine experiment which has an energy threshold an
order-of-magnitude lower).

Can one understand why the Monte Carlo simulations produce such
well-defined theoretical predictions?  Yes, there are at least five
reasons, which I list below in what I judge to be the
relative order of
importance.  1) {\it The luminosity boundary condition} requires that the
computed photon luminosity of the present-day solar model equals the
measured solar luminosity,$L_{\odot}$,
which is known experimentally to an accuracy
of about two parts in a thousand.  If one oversimplifies the problem of
stellar evolution and represents the output of a solar model in terms of
just the central temperature, $T_c$ (as is done in several recent
papers by different authors), then~\cite{Bahcall89}
the flux of the most-sensitive neutrino branch, the
\b8 neutrinos is $\phi(^8{\rm B}) \propto T_c^{18}$ and
the luminosity $L_{\odot} \propto T_c^{4}$.  One concludes by this argument
that the uncertainty in the \b8 neutrino flux is very small,
\begin{equation}
\left({ {\Delta \phi(^8{\rm B})} \over {\phi(^8{\rm B})} }\right)
{}~=~\left({18 \over 4}\right) \times
\left({{\Delta L_{\odot} } \over {L_{\odot}} }\right) ~<~ 0.01 .
\label{eq:deltaphi}
\end{equation}
This argument suggests that the uncertainty in the \b8 neutrino flux is
less than 1\%.
Actually, the uncertainty I
estimate is very much larger, 14\% .
The reason for the discrepancy between
Eq.~(\ref{eq:deltaphi})
and the uncertainty obtained from a detailed analysis~\cite{Bahcall92}
is that the representation of a
solar model in terms of just a central temperature is a gross
oversimplification.
(The computed neutrino flux is an integration of the local production
rate over the temperature-density profile of the model sun and also
depends, for example,
in different ways upon the different input nuclear cross sections.)
Nevertheless, you can see by this argument that the luminosity
boundary condition provides a severe constraint on the allowed values
of the neutrino fluxes.
2) {\it The precision of the input parameters}
has greatly improved over the years as
many individuals
and groups (physicists, chemists, and astronomers) have remeasured and
recalculated the quantities required to determine the solar
model neutrino fluxes.
The recently-evaluated uncertainties are relatively small, in large part,
because of this successful community effort.
3) {\it Helioseismologists} have measured the frequencies of
thousands of solar pressure modes to an accuracy of better than one part
in a thousand.  The standard solar modes used to calculate solar
neutrino fluxes reproduce the measured $p-mode$ eigenfrequencies to
typically one part in a thousand, establishing the basic correctness of
the solar model to a depth of at least half the solar radius.
One no longer has the freedom to speculate about radically
different possible solar models because of the many precisely measured
helioseismological frequencies.
4) {\it The sun is in a simple state of stellar
evolution}, the main sequence,
and we know more about it experimentally than about any other
star.  The physics of the interior of the sun is relatively
simple; for example, detailed corrections to the equation of state are
only of order of a few percent.
5) {\it There are many input parameters},
including the cross sections for all of the relevant nuclear reactions,
the solar luminosity, and the surface heavy element abundances.  In any
particular time period, the improvements in some of these parameters cause
the calculated neutrino event rates to increase and the improvements
in other parameters
cause the calculated neutrino event rates to decrease.  On the average,
the best-estimate for the chlorine experiment has remained within a
narrow range  over the past 25 years (see
Figure~1.2 of Ref.~\cite{Bahcall89}).  If we consider all of
my published calculations
in which a full evaluation was made including an estimated theoretical
error, then the range over the last quarter century has been between
$5.8 $ SNU and $10.5$ SNU, the midpoint of which is within 0.3 SNU of
the current best estimate.

\section{Direct Comparison of Chlorine and\hfil\break Electron-Scattering
Experiments}
\label{sec-DirComparison}

\noindent
The chlorine and the Kamiokande experiments are sensitive, to a large extent,
to the same neutrino source, the rare \b8 neutrinos.  The Kamiokande
experiment measures only \b8 neutrinos.
For the chlorine experiment,  about 78\% of the standard-model
calculated rate is from the same source.
The chlorine and the Kamiokande experiments differ in that the threshold
for chlorine (0.8 MeV) is about an order-of-magnitude larger than for
Kamiokande (7.5 MeV).

We will compare directly the results for these two experiments using a
lemma, proved in Ref.~\cite{Bahcall91}, that states that
the shape of the \b8 neutrino
spectrum that is produced in the center of the sun is the same,
to an accuracy of one part in $10^5$, as the
shape of the known spectrum that is produced in terrestrial
laboratories.  The largest imprints
of the solar environment are caused by Doppler shifts and by the
gravitational redshift, but both of these effects are negligibly small
for our purposes.  Therefore, the shape of the neutrino spectrum must be
the same in a terrestrial laboratory and in the center of the sun unless
physics beyond the standard electroweak model causes energy-dependent
changes in the neutrino spectrum.

We know from the Kamiokande experiment how many \b8 neutrinos reach the
earth with energies about 7.5 MeV.  If standard electroweak theory is
correct, then we can extend the laboratory \b8 spectrum, normalized by
the Kamiokande results, down to 0.8 MeV, the threshold for the chlorine
experiment.  This leads to a minimum predicted rate for the chlorine
experiment based on scaling the Kamiokande results down to the chlorine
threshold and on ignoring all other neutrino sources except the rare
\b8 neutrinos.  This minimum value is

\begin{equation}
{\rm Cl~Rate~(^8B ~ only)} ~=~ {\rm {\left({Rate~Observed}
\over {Rate~Predicted}\right)_{Kamiokande} }}
\left(^8{\rm B}\right)
\times 6.2~SNU,
\label{eq:minimum}
\end{equation}
or

\begin{equation}
{\rm Cl~Rate~(^8B ~ only)} ~\geq~ 3.1~SNU ~>~2.2~SNU.
\label{eq:Rateminimum}
\end{equation}
In Eq.~(\ref{eq:minimum}), 6.2 SNU is the capture rate for
chlorine that is predicted by the
standard model for just the \b8 neutrinos.  The result shown in
Eq.~(\ref{eq:Rateminimum}) indicates that the flux of just \b8 neutrinos
that are seen in the
Kamiokande~II experiment is by itself sufficient to yield a capture rate
in excess of the chlorine experimental value of $2.28 \pm 0.23 $ SNU.  The
additional neutrinos from other, more reliably calculated  branches of the
$pp$ fusion chain, further increase the discrepancy.

What is the most serious mistake that we could have made in the solar
model calculations?  The most crucial error would have been to have
calculated wrongly the \b8 neutrino flux since only \b8 neutrinos are
observed in the Kamiokande experiment and \b8 neutrinos also account for
nearly 80\% of what is expected
in the chlorine experiment.  Suppose that this flux was calculated
wrongly, perhaps because all of the laboratory nuclear physics
measurements of the reaction that produces \b8 have been seriously in error.
Would it then be possible to reconcile the chlorine and the Kamiokande
experiments?

The answer to this question is given in Figure~4 and is ``No''.
For each of the 1000
solar models discussed earlier, I have replaced the calculated \b8 flux
by a value drawn from a normal distribution with the mean and the
standard deviation determined by the Kamiokande experiment.  This
assumption reduces {\it ad~hoc} the mean rate by about 3.1 SNU, as
indicated by Eq.~(\ref{eq:Rateminimum}).  The resulting histogram is now
centered just below 5 SNU, instead of at 8 SNU, as in the unfudged original
calculations (see Figure~2).  In addition, the width of the histogram is
much narrower than in the actual calculations because the
contribution of the \b8 neutrinos is reduced and \b8 neutrinos are the most
uncertain of all the solar neutrino sources.

Even in the worst case scenario shown in Figure~4, in which the
normalization of the \b8 neutrino flux is artificially adjusted to equal
the measured Kamiokande~II value, the calculated rate for the chlorine
experiment is
many experimental standard deviations larger than the observed rate.
Hans Bethe and I have concluded~\cite{BahcallBethe93}
on the basis of Figure~4 that either new physics (beyond the standard
electroweak model) is required to change the shape of the \b8 neutrino
energy spectrum or one of the two experiments (chlorine and
Kamiokande~II) is wrong.

\section{The Gallium Experiments: Further Evidence}
\label{sec-galliumexpts}

\noindent
More than half (54\%), or $71~$ SNU,
of the predicted standard model event rate, $132_{-6}^{+7}~$SNU, in the
gallium experiments comes from the low-energy $pp$ neutrinos.
The standard flux of
these  neutrinos can be calculated with precision (accuracy exceeding
$1$\%).  They are not observable with any of the other
currently-operating experiments (or even other funded experiments
under development).  The $^7$Be neutrinos, which can be calculated with
moderately high precision (6\%), also contribute significantly to the
predicted standard capture rate, $36~$SNU, or 27\% of the total gallium rate.
The \b8 neutrinos, which dominate---according to the standard model---the
chlorine and the
Kamiokande~II experiments, contribute less than 10\% to the standard
theoretical rate.

As shown in Figure~1, the capture rates measured in the
GALLEX and the SAGE solar neutrino experiments
are both about 2.9 SNU below the standard model predictions.
These results strengthen the conclusion that new physics is required to
explain the solar neutrino problem.  Since the gallium experiments are
most sensitive to low energy neutrinos and the chlorine and
Kamiokande~II experiments are most sensitive to higher-energy neutrinos,
the results from the SAGE and GALLEX experiments cannot be compared
directly with the chlorine or the Kamiokande~II experiments without
introducing a specific theoretical model.

\section{Which New Physics?}
\label{sec-newphysics}

\noindent
The two most popular mechanisms for explaining the solar neutrino
problem via new physics are vacuum neutrino oscillations, first
discussed in this connection by Gribov and Pontecorvo~\cite{Gribov69} in an
epochal paper, and matter-enhanced neutrino oscillations,
the MSW effect, a beautiful
idea discovered by
Wolfenstein~\cite{Wolfenstein78} and by Mikheyev and
Smirnov~\cite{Mikheyev86}.
Other solutions have been proposed for the solar neutrino problem that involve
new weak interaction physics.  These other solutions include
rotation of the neutrino magnetic moment~\cite{Cisneros71}, matter-enhanced
magnetic moment transitions~\cite{Lim88}, and neutrino decay~\cite{Bahcall72}.

If new physics is required, then the MSW effect is in my view the most
likely candidate.  Non-zero neutrino masses and mixing angles are required
for the MSW effect to occur in a plausible way, but the indicated masses
and mixing angles are within the range that is expected on the basis of
Grand Unified Theories.  The MSW effect can work without fine tuning and
with a natural extension of the simplest version of the
standard electroweak model.  If the MSW effect is the explanation of the
solar neutrino problem, then the chlorine, gallium, and Kamiokande
experimental results (summarized in Figure~1) imply that at least one
neutrino coupled to the electron-flavor neutrino has mass and mixing
angle that satisfy~\cite{Krastev93}: $\delta m^2 ~\sim~ 10^{-5} {\rm ~eV^2}$
and $sin^2{2\theta} ~\sim~10^{-2}$ or
$\delta m^2 ~\sim~ 10^{-5} {\rm ~eV^2}$
and $sin^2{2\theta} ~\sim~0.6$.

\section{New Experiments}
\label{sec-newexperiments}

\noindent
Table~2 describes the five new solar neutrino experiments that are
funded for operation or for development.
Each of the modes of each of the experiments
listed in Table~2 is expected to yield more than 3,000 neutrino events
per year (except for the $\nu - e$ scattering mode of SNO).
In one year, each experiment will record
more than three times the total number of
neutrino events that have been counted to date in all solar neutrino
experiments since the chlorine experiment began operating a quarter of a
century ago.  With this greater statistical accuracy,
solar neutrino physics will become a more precise subject.

\begin{table}[t]
\begin{center}
\begin{tabular}{lcc@{\hspace{0pt}}l}
\multicolumn{4}{c}{\hfil Table 2.\ New Solar Neutrino
Observatories\hfil}\\
\noalign{\smallskip}
\multicolumn{4}{c}{\hfil Typical Event Rates $\gtorder~3 \times
10^3$~{\rm yr}$^{-1}$\hfil}\\
\noalign{\medskip\hrule\medskip}
\hfil Observatory\hfil&$E_{Th}(\nu)$
&\multicolumn{2}{c}{\hfil Reaction(s)\hfil}\\
&(MeV)\\
\noalign{\medskip\hrule\smallskip\hrule\medskip}
SNO&6.4&$\nu_e + {\rm ^2H} \to$&$ p + p + e^-$\\
\noalign{\medskip}
&2.2&$\nu + {\rm ^2H} \to$&$ n + p + \nu$\\
\noalign{\medskip}
&5&$\nu + e^- \to$&$ \nu + e^-$\\
\noalign{\bigskip\smallskip\medskip}
Super-&5&$\nu + e^- \to$&$ \nu + e^-$\\
Kamiokande\\
\noalign{\bigskip\smallskip\medskip}
ICARUS&$\sim 10$&$\nu_e + {\rm ^{40}Ar} \to$&$ e^- + {\rm ^{40}K}^*$\\
\noalign{\medskip}
&5&$\nu + e^- \to$&$ \nu + e^-$\\
\noalign{\bigskip\smallskip\medskip}
BOREXINO&0.4&$\nu (^7{\rm Be})+ e^- \to$&$ \nu (^7{\rm Be}) + e^-$\\
\noalign{\bigskip\smallskip\medskip}
HELLAZ&0.1&$\nu + e^- \to$&$ \nu + e^-$\\
\noalign{\medskip\hrule}
\end{tabular}
\end{center}
\end{table}

The experiments are listed in order of their expected completion
dates:\ SNO (1996~\cite{Aardsma87}),
Superkamiokande (1996; see Ref.~\cite{Takita93,Totsuka90},
BOREXINO\break
($\geq$~1996; \cite{Raghavan90}),
ICARUS (1998; see Ref.~\cite{Revol93}),
and HELLAZ (proposed, not yet approved~\cite{Seguinot92}).
Table~2 lists the neutrino
threshold energy for
each reaction mode and the individual reactions that will be observed.
I have not listed other promising experimental proposals because it is not
yet clear which of these possibilities will receive funding.
In particular, a prototype detector of $pp$ neutrinos making use of the
properties of superfluid helium has been tested successfully and
appears to be feasible~\cite{Lanou87,Porter94}
It is
clear, however, that the experiments listed in Table~2 will be insufficient to
uniquely solve for all of the fundamental neutrino parameters. Other
experiments are required to establish uniqueness in the inferences and
to provide a measure of redundancy to assure ourselves that systematic
experimental uncertainties have not misled us.

The SNO experiment has two
capabilities for testing, independent of solar models, the inference
that physics beyond the standard model is required.  They are:\ 1) SNO will
measure the energy
spectrum of electron-flavor neutrinos
above 5~MeV in the charged current reaction (neutrino absorption
by deuterium)
and 2) SNO will measure the total neutrino flux independent of flavor in the
neutral current reaction (neutrino disintegration of deuterium).

As emphasized earlier, the shape of the \b8 neutrino energy spectrum
is independent of solar-model uncertainties.
A measurement of the
neutrino energy spectrum could establish that physics beyond the
standard electroweak model is required.

The comparison of the neutrino
fluxes measured via neutrino absorption on deuterium
and by neutrino disintegration of deuterium will
test the equality of the charged and the neutral currents.
If the total neutrino flux is not equal to the electron neutrino flux,
this would be direct evidence for neutrino flavor changing.
The charged and neutral currents must be
equal unless some neutrinos change their flavor after they are created
in the solar core.  Unfortunately, no energy information will be
available for the neutral current detection.  Also, the neutral and
charged-current fluxes would be equal even if some of the original
electron-type neutrinos changed into sterile neutrinos.

Like SNO, ICARUS can measure the shape of the \b8 neutrino energy
spectrum via neutrino absorption.  Moreover, ICARUS has a unique
``smoking-gun'' signal for neutrino absorption, the $\gamma$ decay of
the excited state of $^{40}$K.

There will be welcome redundancy if all experiments
operate as planned.  Three experiments
(Superkamiokande, SNO, and ICARUS) will measure for \b8
neutrinos the $\nu_e - e$
scattering rate and the recoil electron energy spectrum; the electron
recoil spectrum reflects the incoming neutrino energy spectrum.
The fact that the Superkamiokande experiment contains more than 30 times
the fiducial volume for solar neutrino experiments
as the highly-productive Kamiokande experiment is an
indication of the amount of improvement that may be expected in the next
generation of solar neutrino experiments compared to those performed to
date.

The BOREXINO and HELLAZ experiments are essential
in order to distinguish between different new-physics possibilities.
These experiments are the only ones currently
under development that will measure the energy of
individual events with energies less than 5~MeV.
The threshold for BOREXINO is 0.4 MeV and for HELLAZ is 0.1 MeV.
These experiments must be performed
in order to determine the neutrino survival probability at low
energies. The BOREXINO and HELLAZ experiments also
have another highly desirable feature; they will both
measure the $\nu_e$ flux at a
specific energy, the energy (0.86 MeV) of the \be7 neutrino line.
The theoretical predictions are more specific, and therefore the
measurements are more diagnostic, when the neutrino flux at a specific
energy is observed.

The HELLAZ experiment is unique among the experiments being developed;
it is the only experiment being developed
to observe individual events from the basic $pp$ reaction
(maximum energy 0.4 MeV).  In addition, HELLAZ has the energy resolution
to potentially measure the predicted~\cite{Bahcall93} 1.29 keV shift
between the average energy of the solar \be7 line and the laboratory
energy of the line.  A measurement of this energy shift, which is due to
thermal effects in the center of the sun, is equivalent to a direct
measurement of the central temperature of the sun.

\section{Conclusions}
\label{sec-conclusions}

\noindent
The field of solar neutrino research is flourishing.  The four
operating
experiments have confirmed that the sun shines via nuclear fusion
reactions that produce MeV neutrinos (see Eq.~(\ref{eq:Heburning}) ).
There are differences between the predictions and the observations
(see Figure~1), but these differences are within the usual range of
astronomical uncertainties (generally a factor of two or three).  The
agreement between theory and observation is, from the astronomical point
of view, remarkably good because the calculated neutrino fluxes depend
sensitively upon the interior conditions.

Nevertheless, all four experiments disagree with the corresponding
theoretical predictions based upon the simplest version of the standard
electroweak theory.  These disagreements are larger than the estimated
uncertainties.  The luminosity boundary condition and the
helioseismological measurements are especially important in guaranteeing
the robustness of the theoretical predictions
(see discussion in \ref{sec-Comparison}).  Monte Carlo experiments that
make use of 1000 implementations of the standard solar model indicate
that the chlorine and the Kamiokande~II (water-Cerenkov) experiments
cannot be reconciled without an energy-dependent change in the \b8
solar neutrino spectrum relative to the laboratory spectrum (see
Figure~2-Figure~4).  New
physics is required to explain an energy-dependent change in the shape
of the neutrino spectrum.
The gallium experiments, GALLEX and SAGE, strengthen the conclusion that
new physics is required.

New experiments, SNO, Superkamiokande, and ICARUS,
will test the conclusion that new physics is required independent of
uncertainties due to solar models.  These experiments can determine the
shape of the \b8 solar neutrino energy spectrum and whether or not
electron-flavor neutrinos have oscillated into some other flavor
neutrinos.

\clearpage
\noindent
{\Large\bf Figure Captions}

Figure 1.\ Comparison of
measured
rates~\cite{Davis93,Hirata89,Hirata91,Suzuki93,Anselmann92,
Anselmann93,Abazov91a,Abazov91b,Bowles93}
and standard-model predictions~\cite{Bahcall92} for four solar-neutrino
experiments.

Figure 2.\ 1000 solar
models vs
experiments\cite{BahcallBethe93}.  The number of precisely
calculated solar models that
predict different solar neutrino event rates are shown for the chlorine
experiment\cite{Davis93}.
Each input parameter in each solar model was drawn
independently from a normal distribution having the mean and the
standard deviation appropriate to that parameter.  The experimental
error bar includes only statistical errors (1$\sigma$).

Figure 3.\ 1000 solar
models vs
experiments~\cite{BahcallBethe93}.
The number of precisely calculated solar models that
predict different solar neutrino event rates are shown for the
Kamiokande experiment~\cite{Hirata89,Hirata91}.
The solar models from which the fluxes were derived satisfy
the equations of stellar evolution including the boundary conditions
that the model luminosity, chemical composition, and effective
temperature at the current solar age be equal to the observed values.
Each input parameter in each solar model was drawn
independently from a normal distribution having the mean and the
standard deviation appropriate to that parameter.  The experimental
error bar includes only statistical errors (1$\sigma$).

Figure 4.\ 1000
artificially modified fluxes~\cite{BahcallBethe93}.
The $^8$B neutrino fluxes computed for
the 1000 accurate solar models were replaced in the figure shown by
values drawn randomly for each model from a normal distribution with the
mean and the standard deviation measured by the Kamiokande experiment
\cite{Hirata89,Hirata91}.

\begin{thebibliography}{}
\itemsep=0pt
\bibitem{Schatzman51}E. Schatzman, Comptes Rendus {\bf 232}, 1740
(1951).
\bibitem{Bahcall64}J. N. Bahcall, Phys. Rev. Lett. {\bf 12}, 300 (1964).
\bibitem{Davis64}R. Davis Jr., Phys. Rev. Lett. {\bf 12}, 303 (1964).
\bibitem{Davis93}R. Davis Jr., \frontiers, p.~47.
\bibitem{Hirata89}K. S. Hirata, et al., Phys. Rev. Lett. {\bf 63}, 16
(1989).
\bibitem{Hirata91}K. S. Hirata, et al., Phys. Rev. D. {\bf 44}, 2241
(1991).
\bibitem{Suzuki93}Y. Suzuki, \frontiers, p.~61.
\bibitem{Anselmann92}P. Anselmann, et al., Phys. Lett B {\bf 285},
376 (1992).
\bibitem{Anselmann93}P. Anselmann  Phys. Lett. B {\bf 314}, 445 (1993).
\bibitem{Abazov91a}A. I. Abazov, et al., Nucl. Phys. B (Proc. Suppl.)
{\bf 19}, 84 (1991a).
\bibitem{Abazov91b}A. I. Abazov, et al, Phys. Rev. Lett. {\bf 67} 3332
(1991b).
\bibitem{Bowles93}T. Bowles and V. N. Gavrin, private communication
(1993).
\bibitem{Bahcall92}J. N. Bahcall and M. H. Pinsonneault, Rev. Mod. Phys.
{\bf 64}, 885 (1992).
\bibitem{Bahcall89}J. N. Bahcall, {\it Neutrino Astrophysics} (Cambridge
University Press, Cambridge, England, 1989).
\bibitem{Bahcall91}J. N. Bahcall, Phys. Rev. D {\bf 44}, 1644 (1991).
\bibitem{BahcallBethe93}J. N. Bahcall and H. A. Bethe, Phys. Rev. D {\bf 47},
 1298 (1993).
\bibitem{Gribov69}V. Gribov and B. Pontecorvo, Phys. Lett. B {\bf 28},
493 (1969); S. M. Bilenky and B. Pontecorvo, Phys. Rep. {\bf 41}, 225
(1978); S. M. Bilenky and S. T. Petcov, Rev. Mod. Phys. {\bf 59}, 671
(1987). See also B.~Pontecorvo, Sov. Phys. JETP {\bf 26}, 984 (1968).
\bibitem{Wolfenstein78}L. Wolfenstein, Phys. Rev. D {\bf 17}, 2369
(1978); Phys. Rev. D {\bf 20}, 2634 (1979).
\bibitem{Mikheyev86}S. P. Mikheyev and A. Yu. Smirnov, Sov. J. Nucl.
Phys. {\bf 42}, 913 (1986); Sov. Phys. JETP {\bf 64}, 4 (1986); Nuovo
Cimento {\bf 9C}, 17 (1986); T. K. Kuo and J. Pantaleone, Rev. Mod.
Phys. {\bf 61}, 937 (1989); S. P. Mikheyev and A. Yu. Smirnov, Progress in
Part. and Nucl. Physics {\bf 23}, 41 (1989).
\bibitem{Cisneros71}A. Cisneros, Astrophys. Space Sci. {\bf 10},
87 (1971); J. Schechter
and J. W. F. Valle, Phys. Rev. D {\bf 24}, 1883 (1981); M. B. Voloshin,
M. I. Vysotsky, and L. B. Okun, Sov. Phys. JETP {\bf 64}, 446 (1986);
M. B. Voloshin and M. I. Vysotsky, Sov. J. Nucl. Phys.
{\bf 44}, 544 (1986); L. B. Okun, Sov. J. Nucl. Phys. {\bf 44}, 546
(1986).
\bibitem{Lim88}C. S. Lim and W. J. Marciano, Phys. Rev. D {\bf 37}, 1368
(1988); E. Kh. Akhmedov, Phys. Lett. B {\bf 213}, 64 (1988).
\bibitem{Bahcall72}J. N. Bahcall, N. Cabibbo, and A. Yahil, Phys. Rev.
Lett. {\bf 28}, 316 (1972); S. Pakvasa and K. Tennakone, Phys. Rev.
Lett. {\bf 28}, 1415 (1972); Z. G. Berezhiani and M. I. Vysotsky, Phys.
Lett. B {\bf 199}, 281 (1987); J. Frieman, H. Haber, and K. Freese,
Phys. Lett. B {\bf 200}, 115 (1988); X.-G. He, S. Pakvasa, and R. S. Raghavan,
Phys. Rev. D {\bf 38}, 1317 (1988); A. Acker, A. Joshipura, and S.
Pakvasa, Phys. Lett. B {\bf 285}, 371 (1992); Z. G. Berezhiani, G.
Fiorentini, M. Moretti, and A. Rossi, Z. Phys. C---Particles and Fields
{\bf 54}, 581 (1992).
\bibitem{Krastev93}P. I. Krastev and S. T. Petcov, Phys. Lett.B {\bf
299}, 99 (1993); L. Krauss, E. Gates, and M. White, Phys. Lett. B {\bf
299}, 94 (1993); X. Shi, D. N. Schramm, and J. N. Bahcall, Phys. Rev.
Lett. {\bf 69}, 717 (1992); N. Hata and P. Langacker, Phys. Rev. D
(subbmited, 1993).
\bibitem{Aardsma87}G. Aardsma {\it et al.}, Phys. Lett. B {\bf 194}, 321
(1987); G. Ewan et al., Sudbury Neutrino Observatory Proposal
SNO-87-12 (1987);
H. Chen, Phys. Rev. Lett. {\bf 55}, 1534 (1985).G. Ewan et al., Sudbury
Neutrino Observatory Proposal
SNO-87-12 (1987); H. Chen, Phys. Rev. Lett. {\bf 55}, 1534 (1985).
\bibitem{Takita93}M. Takita, \frontiers, p.~135.
\bibitem{Totsuka90}Y. Totsuka, in {\it Proceedings of
the International
Symposium on Underground Physics Experiments}, edited by K. Nakamura
(ICRR, University of Tokyo, 1990), p.~129;
A. Suzuki, in {\it
Proceedings of the Workshop on Elementary-Particle Picture of the
Universe}, Tsukuba, Japan, 1987, edited by M. Yoshimura, Y. Totsuka, and
K. Nakamura (KEK Report No. 87-1, Tsukuba, 1987), p.~136.
\bibitem{Raghavan90}R. S. Raghavan, in {\it Proceedings of the XXVth
International Conference on High Energy Physics}, Singapore, 1990,
edited by K. K. Phua and Y. Yamaguchi (World Scientific, Singapore,
1990), Vol. 1, p.~482;
G. Ranucci for the Borexino Collaboration, Nucl. Phys. B (Proc. Suppl.)
32,
149 (1993);
C. Arpasella {\it et al.}, in ``Borexino at Gran
Sasso: Proposal for a real-time detector for low energy solar
neutrinos," Vols. I and II, University of Milan, INFN report
(unpublished).
\bibitem{Revol93}J. P. Revol,
\frontiers, p.~167; J. N. Bahcall, M. Baldo-Ceolin, D.
Cline, and C. Rubbia,
Phys. Lett. B {\bf 178}, 324 (1986); C. Rubbia, CERN-EP Internal Report
77-8 (1977).
\bibitem{Seguinot92}J. Seguinot, T. Ypsilantis, and A. Zichini, ``A High
Rate Solar Neutrino Detector with Energy Determination," LPC92-31,
College de France, 12/8/92.
\bibitem{Lanou87}R. E. Lanou, H. J. Maris, and G. M. Seidel, Phys. Rev.
Lett.
{\bf 65} , 1297
(1987).
\bibitem{Porter94}F. S. Porter, Phd. Thesis, Brown University
(1994,unpublished).
\bibitem{Bahcall93}J. N. Bahcall, Phys. Rev. Lett. {\bf 71} (15), 2369
(1993).
\end{thebibliography}
\end{document}